\pdfoutput=1
\documentclass[aps, groupedaddress, showpacs, showkeys, 11pt]{revtex4}
\usepackage{newcent, amssymb, amsbsy, amsmath, latexsym, dsfont, array, layout, graphicx, mathrsfs}
\usepackage{color}
\usepackage{appendix}
\newcommand{\fig}[1]{Fig.~\ref{#1}}

\begin{document}

\title{Theory of non-equilibrium single  electron  dynamics  in  STM  imaging  of  dangling  bonds  on a  hydrogenated  silicon  surface}
\author{Lucian Livadaru}
\email[]{lucian@ualberta.ca}
\author{Jason Pitters}
\author{Marco Taucer}
\author{Robert A. Wolkow}

\address{Department of Physics, University of Alberta \& National
Institute for Nanotechnology, National Research Council, Edmonton,
AB T6G 2M9, Canada}


\begin{abstract} 
During fabrication and scanning-tunneling-microscope (STM) imaging of dangling bonds (DBs) on a hydrogenated silicon surface, we consistently observed halo-like features around isolated DBs for specific imaging conditions. These surround individual or small groups of DBs, have abnormally sharp edges, and cannot be explained by conventional STM theory. Here we investigate the nature of these features by a comprehensive 3-dimensional model of elastic and inelastic charge transfer in the vicinity of a DB. Our essential finding is that non-equilibrium current through the localized electronic state of a DB determines the charging state of the DB. This localized charge distorts the electronic bands of the silicon sample, which in turn affects the STM current in that vicinity causing the halo effect. The influence of various imaging conditions and characteristics of the sample on STM images of DBs is also investigated. 
\end{abstract}

\pacs{6116Ch, 7220Jv, 7320-r, 7323Hk}

\maketitle

\section{Introduction}
Dangling bonds (DBs) on silicon crystal surfaces are becoming increasingly important candidates for building extremely miniaturized nanoelectronics devices.
DBs are band gap states that can act as donors or acceptors, depending on their energy level with respect to the Fermi level. A surface silicon atom can host a dangling bond when only three of its valence electrons are involved in covalent bonds, the remaining valence electron being left in an unsatisfied (dangling) bond state. It turns out that a DB can be in three, energetically different states: a negative state $DB^-$ (holding two localized electrons); a neutral state $DB^0$ (holding a single localized electron); and positive state $DB^+$ (holding a localized hole). 

The exact energy levels for an isolated DB on an otherwise perfectly terminated H:Si (001) 2x1 surface are not exactly known, but reasonable estimates can be obtained by various computational methods: an extended Huckel theory \cite{raza2007theoretical}, Poisson-Schroedinger equations \cite{blomquist2006controlling}, and density functional theoretical (DFT) methods\cite{haider1}. The three-dimensional form of the DB wavefunction must, as a surface state, decay exponentially into vacuum and the crystal, as well as laterally away from the center of the host Si atom. Theoretical estimates~\cite{haider1,raza2007theoretical,blomquist2006controlling} and experimental measurements (by contactless capacitance-voltage method\cite{yoshida2000ultrahigh})  of the DB-state energy in the bandgap are found to agree relatively well. In this study we assume the energy levels are, with respect to the silicon valence band maximum (VBM): $E_{\rm DB-}$= 0.82 eV~\cite{yoshida2000ultrahigh}, $E_{\rm DB0}$= 0.35 eV~\cite{gaussian}, $E_{\rm DB+}$= 0.0 eV~\cite{blomquist2006controlling}. Because the latter state is degenerate with the silicon valence band, in this study it is assumed not to play a significant role.

Studying constant-current empty-state scanning tunneling microscope (STM) images for different sample bias voltage, $V_{\rm bias}$, and current setpoints reveals in some cases the existence of a dark halo surrounding a bright spot of atomic size at the location of the DB. 
The appearance of such a halo has been also observed when imaging single dopant atoms located near the surface for other types of semiconductor~\cite{teichmann2008controlled, lee2010tunable}. 
The simple and intuitive explanation is that this dark halo is caused by upward bending of the energy bands at the surface, leading to a reduction of the STM current (in constant-height imaging mode) or equivalently a decrease in the STM tip height (in constant-current imaging mode). 
In order to distinguish such spatially limited band bending from one of a large scale, e.g. for a great density of DBs, it is important to understand to what extent band bending due to the charging of a collective of such surface states occurs \textit{uniformly} along the surface and where the delimiting behavior occurs. Arguably, a rule of thumb is that band bending becomes uniform (and the Fermi level becomes pinned at the surface) when the average distance between two dangling bonds become equal to the screening length in the semiconductor. However, for the cases approached in this study, due to the great extent of H-termination of our silicon surfaces, this condition is not fulfilled. Therefore, the Fermi level is not pinned at such surfaces.  This fact implies that each isolated DB (unless very close to others), has its own individual screening field, due to mobile charge carriers in the crystal. On the other hand, isolated \textit{groups} of closely spaced DBs ($<$ 2nm) are electronically coupled ~\cite{haider1, raza2007theoretical, pitters2011tunnel}, and have more or less a common screening field around the whole DB group. The coupling of such groups in the presence of an STM tip was recently studied experimentally and theoretically within an extended Hubbard model~\cite{pitters2011tunnel} to yield time-average occupations of individual DBs, without explicitly accounting for the dynamics of electrons in/out the DB group.

The properties of various types of single defect levels in semiconductors have been previously studied in the context of their STM imaging appearance~\cite{ebert2}. For DBs, the carrier capture properties~\cite{berthe2008probing, grandidier2009defect} and the on-site Coulomb interaction between two localized electrons can be extracted from spectroscopic measurements~\cite{nguyen2010coulomb} on isolated DBs on a B-doped Si(111)-$(\sqrt{3}\times\sqrt{3})R30^o$. However in these latter studies, electron dynamics is simpler than for a DB on hydrogenated Si surface, as there is no tip-induced band bending at the surface.
Here we study single electron dynamics for the case when band bending is present at the surface. We account for all the important charge transfer mechanisms in order to gain insight into the charging dynamics of a DB and its appearance in STM images.

\section{DBs in the potential landscape of the imaging probe}
\begin{figure}[tb]
\centerline{
\includegraphics[scale= 0.2, width= 0.9\linewidth]{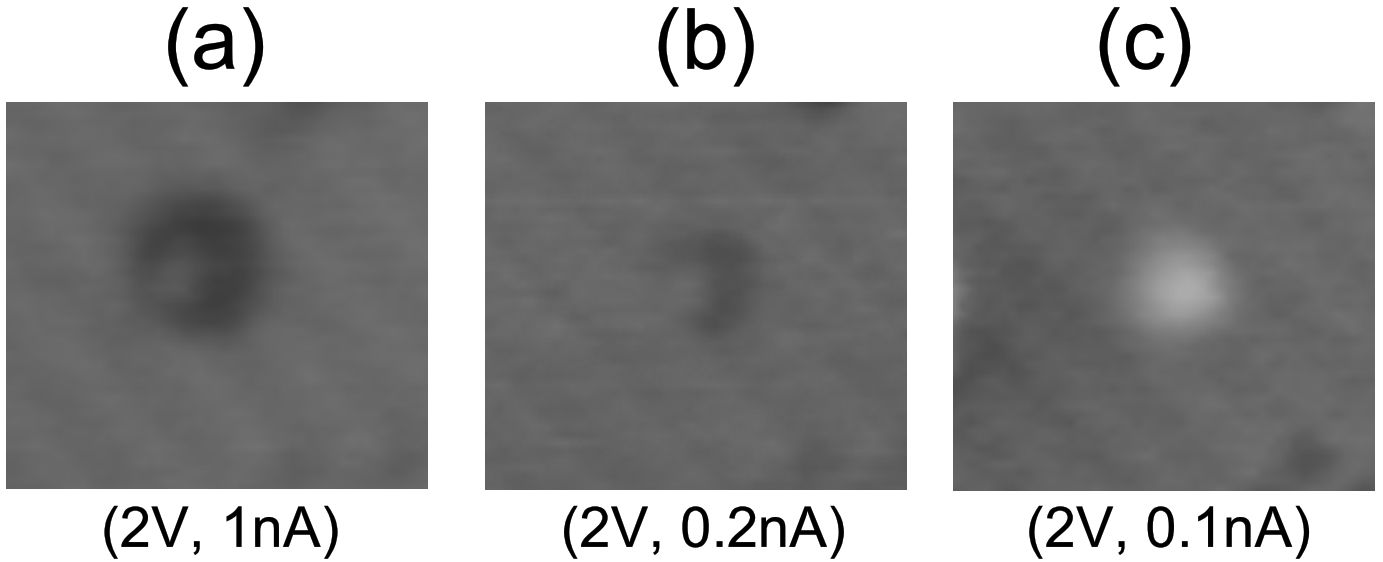}
}
 \caption{STM images of a $4.0\times 3.6$~nm region of the $2\times 1$ H-Si (001) surface with an isolated DB in the unoccupied imaging mode (sample biased positively) at imaging setpoints (sample bias and current) indicated below each image. }
\label{experimImages1}
\end{figure}

In Figure~\ref{experimImages1} we show typical unoccupied-state STM images of dangling bonds on H-terminated Si (001) with a 2$\times$1 surface reconstruction. In part (a) both single and coupled DBs are seen, with halo-like features. The halos around single DBs are consistently more prominent than those around groups of coupled DBs. In (b) a single DB can be identified at the center of the image as a bright spot surrounded by a darker ring or halo-like feature. This appearance is consistent with many experimental observations taken at similar conditions, and has been reported in the literature~\cite{lyding}. Also, such imaging features have been observed in the case of other localized charge centers, such as subsurface dopants~\cite{ebert2}.

\begin{figure}[tb]
\centerline{
 \includegraphics[scale= 0.2, width= 0.9\linewidth]{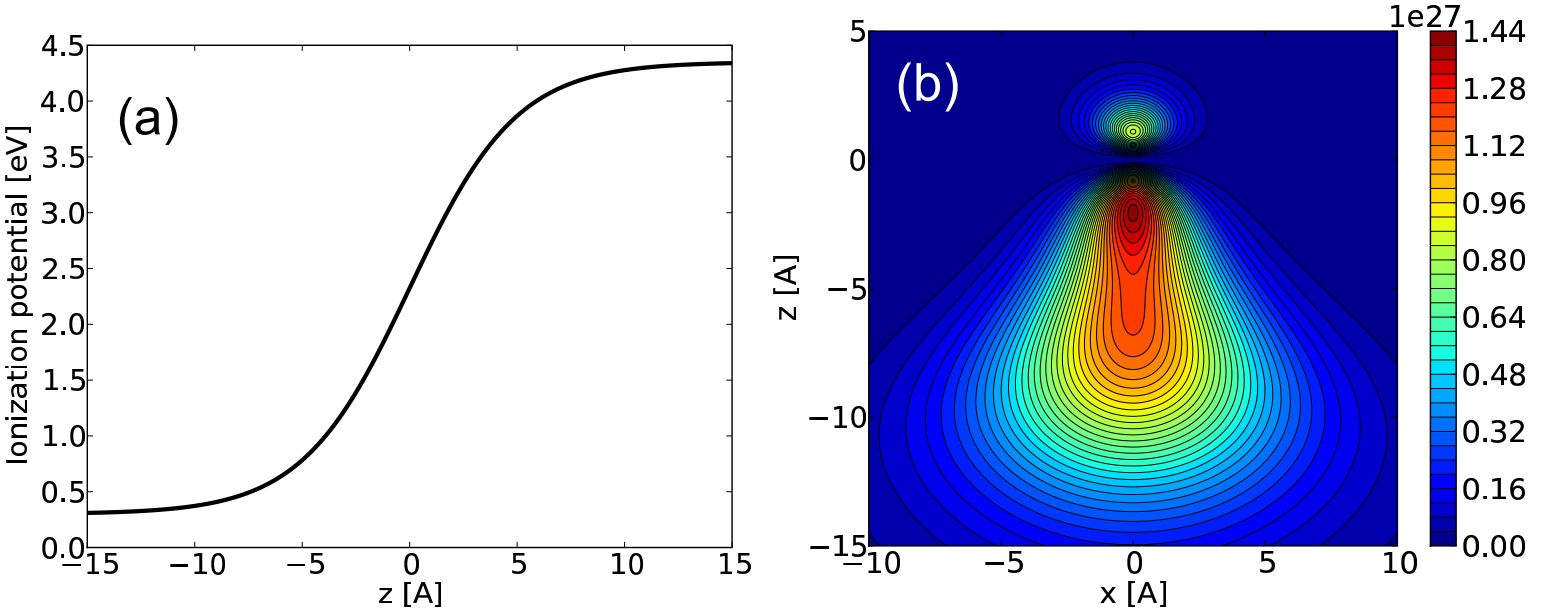}
}
  \caption{Characteristics of the DB orbital. (a) The ionization potential as a function of distance from the surface (positive $z$ in the vacuum). (b) Color map of the distribution of charge density of a DB in an axial plane as predicted by the modified STO function in eq. \ref{STO}. The color bar indicates the values of the density in units of $10^{27}$ m$^{-3}$.}
\label{PsiDB}
\end{figure}

We define an \textit{isolated DB} on the silicon surface as one with no equivalent nearby DBs to be significantly coupled with (nor other resonant electronic levels). This implies that at an isolated DB an electron is strictly localized in an orbital at a single Si atom. We assume a simple Slater-type orbital (STO) for the DB wavefunction having p-type symmetry (two lobes)
\begin{equation}
	\psi_{\rm DB}(\mathbf{r}) = \psi_{\rm DB}(r,\theta) = Nr\cos(\theta) \exp[-\zeta(\theta) r],
	\label{STO}
\end{equation}
where $r$ and $\theta$ are the three-dimensional spherical coordinates and $N$ is a normalization constant.
The difference from a ``regular'' STO is that in our case the decay rate $\zeta$ in the exponential must vary with the $z$-coordinate in order to ensure consistency with the asymptotic decay in the specific environment of the DB, as follows.
The decay rate is related to the ionization potential $W_{\rm i}$ of the DB electron via $\zeta =\sqrt{2mW_{\rm i}}/ \hbar $. This ensures the correct asymptotic behavior (from first-principles) of the wavefunction above. As the DB orbital is located partially in vacuum and partially in silicon, the ionization potential is not a constant, but it rather  depends on the ``ionization path'' of the DB electron, e.g. it is energetically easier to extract an electron toward the bulk (into the conduction band of silicon) than 
toward the vacuum. To reflect this fact, here we assume a simple form for this dependence, namely
\begin{equation}
  W_{\rm i} (z)= W_{\rm bulk} +\frac{1}{2} (W_{\rm vac} -W_{\rm bulk}) [\tanh(z/w) + 1]  
\end{equation} 
where $W_{\rm bulk}$ and $W_{\rm vac}$ are the ionization potentials with respect to bulk and vacuum, respectively, $z$ is measured from the surface and positive toward vacuum, and $w$ is a characteristic width of the transition as can be seen in \fig{PsiDB} (a). The normalization of this wavefunction is calculated numerically and the resulting modified STO function is plotted in \fig{PsiDB} (b) as a color map for a two-dimensional axial section. Note that, as a consequence of our choice for the $\zeta$ function, the two lobes of the DB orbital are quite disproportionate, with the vacuum lobe being much smaller in spatial extent and portion of charge. This is also qualitatively consistent with DB orbital characteristics derived from \textit{ab initio} calculations for an isolated DB at the surface of model hydrogenated silicon clusters\cite{haider1}, mimicking the 2$\times$1 H-Si (001) surface.

When the DB is negatively charged, the localized extra electron is expected to locally distort the electronic levels at the surface of the host silicon crystal, i.e. cause \textit{band bending}. Band bending in a semiconductor is the phenomenon of distortion of electronic levels of the crystal in the presence of an external perturbing electric field. During STM imaging, another source of band bending is the biased STM probe itself and this is known as \textit{tip-induced band bending}. Depending on the value of the STM bias voltage, the amount of band bending can be considerable, sometimes enough to induce an inversion layer near the surface (whenever the Fermi level becomes lower than the intrinsic level of the semiconductor). The amount of tip-induced band bending can be theoretically estimated by solving the Poisson and Schroedinger equations simultaneously, for example by finite element methods (FEM) for an assumed probe geometry. 

\begin{figure}[tb]
\centering
\centerline{
 \includegraphics[scale= 0.2, width= 0.9\linewidth]{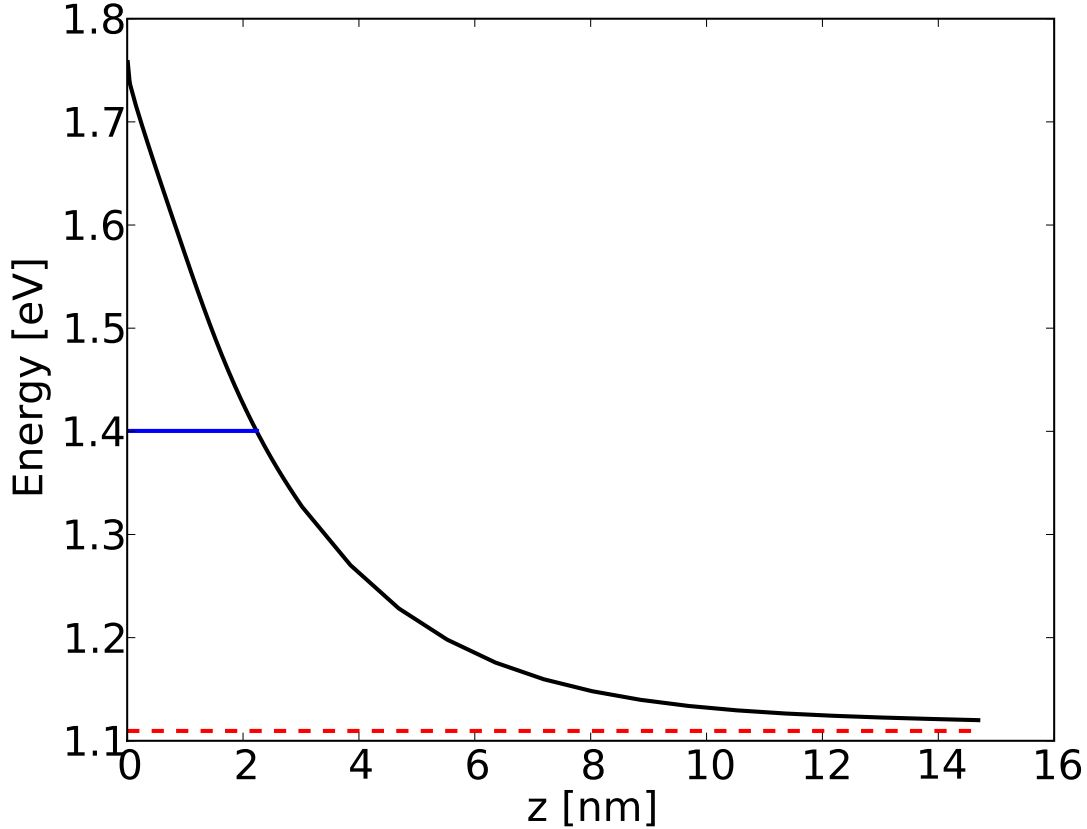}
}  
\caption{Calculated band bending diagram as a function of the $z$-coordinate. The black curve is the CBM, the red dotted line is the sample Fermi level, and the blue line shows the DB$^-$ level with indication to tunneling into resonant levels in the silicon conduction band. The sample is n-type with a resistivity of 3.5 m$\Omega$-cm, sample bias +2 V, STM tip height 7~\AA.}
\label{BBdiagram1}
\end{figure}

During STM imaging of isolated DBs, as the imaging tip is brought closer to the DB location, the band bending is turned gradually up and so is the DB level relative to the sample Fermi level. In Figure~\ref{BBdiagram1}, we show a typical band bending diagram calculated by FEM for the silicon surface in the presence of a scanning probe tip. The sample is n-type with a resistivity of 3.5 m$\Omega$-cm, the sample bias voltage was assumed +2 V, and the tip height was assumed to be 7~\AA. For these typical imaging conditions, the bands are bent upward to the extent that the DB$^-$ level is higher than the sample Fermi level, by 0.3 eV. 
This value is much greater than $kT$, which - assuming thermodynamic equilibrium between the DB state and the bulk semiconductor - means that at room temperature, the DB state should be completely unoccupied. However, the process of STM imaging also inject electrons into the DB state and an interplay of in/out transfer rates starts to form. In this study we show how this fact has important consequences on the way surface states are imaged by STM.

An important consequence is evident from the diagram in Figure~\ref{BBdiagram1}: in the presence of the tip-induced field, the DB level becomes resonant with CB levels of the host crystal, implying that tunneling from DB into the crystal becomes possible. This situation is then similar to double barrier tunneling junctions, which were recently used to study energetic levels of single atoms~\cite{repp2004controlling}, defects~\cite{repp2005scanning}, and molecules~\cite{repp2006scanning,wu2004control,mikaelian2006atomic}, at thin insulating films on metallic surfaces, as well as metallic nano-islands on Si~\cite{oh2002single}. Manipulation of charge states~\cite{repp2006scanning} and bond structure~\cite{mohn2010reversible} of molecules has also been demonstrated. However, the similarity of our system to these other cases is limited because in our case there are a number of other factors at play that contribute to the charging of an individual DB, as we show below.

Careful examination of our available experimental data in conjunction with a quantitative theoretical analysis, indicates that STM imaging features of dangling bonds are due to dangling bond charging caused by \textit{non-equilibrium effects} during imaging, i.e. rate-limited charging/discharging of a DB during extraction/injection of electrons by the STM tip. For routine unoccupied-state imaging conditions (sample bias +2 V, current setpoint 100pA- 1nA) a DB$^-$ state in electrochemical equilibrium with the bulk silicon would be unoccupied under the STM tip, but because the rates of charge transfer are limited, the DB$^-$ state fails to achieve equilibrium and becomes instead a non-equilibrium steady state. STM electrons injected into the DB$^-$ become temporarily trapped there during imaging due to a slow discharge mechanism into the Si sample, which cannot keep up with the rate of injection by the STM tip. 
Explaining the DB imaging as purely quantum mechanical behavior, namely as Friedel oscillations in the local density of states has been proposed - without consideration of band bending and nonequilibrium effects - but this explanation is inconsistent with the variation of the DB features with temperature and STM bias voltage\cite{wenderoth1999low}.

\begin{figure}[tbp]
\centerline{
 \includegraphics[scale= 0.2, width= 0.8\linewidth]{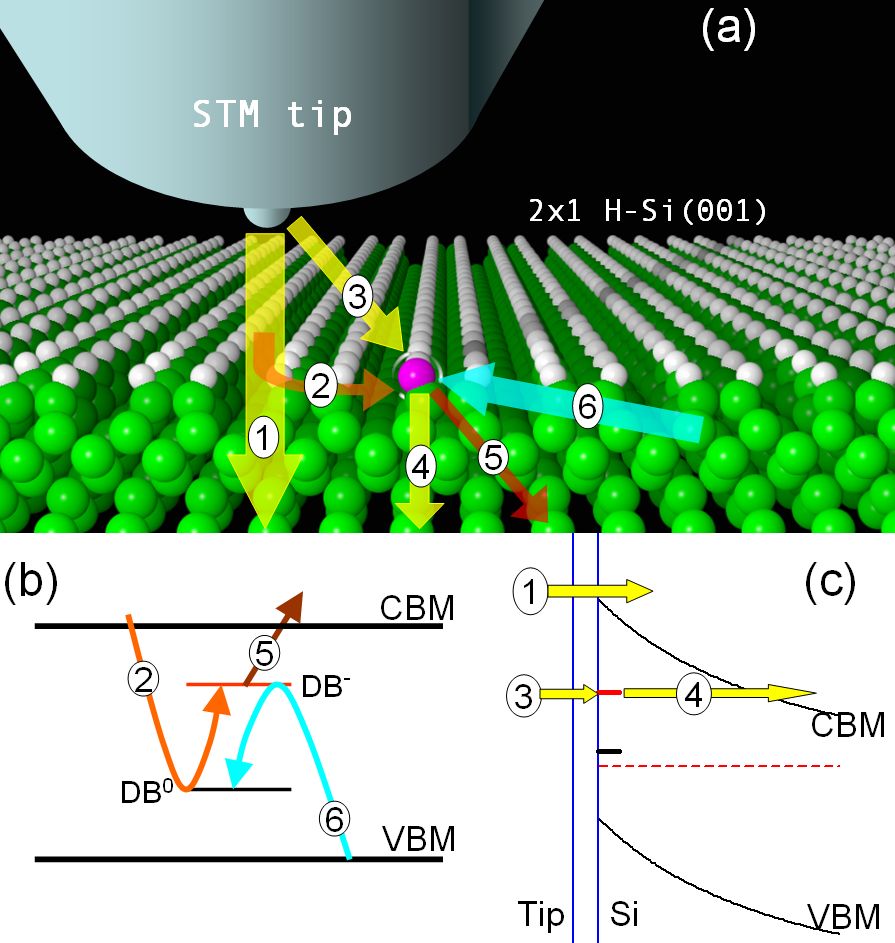}
} 
\caption{Three-dimensional rendering of an isolated DB on the H-Si(001) surface in the presence of an STM tip in the unoccupied imaging mode. (a) Various channels for elastic and inelastic charge transfer between the STM tip and the silicon sample in the vicinity of the DB are indicated by colored arrows and labeled for further discussion. Si atoms are green, H atoms are white and the DB is shown schematically as purple sphere. 
(b) Diagram of the silicon band gap, which further explains the various channels shown in (a): (2) inelastic capture by the DB$^0$ level (black horizontal line) of an STM electron after which DB becomes negative; (5) thermal electron emission from the DB$^-$ level (red horizontal line) into the CB; and (6) inelastic capture of a hole by the negative DB, followed by recombination and the DB becoming neutral.
(c) Tip-induced band bending diagram in the direction normal to the surface. The elastic charge transfers shown in (a) are detailed here: (1) direct tip-silicon tunneling; (3) tip-DB tunneling across the vacuum gap; and (4) tunneling from the DB$^-$ state into the CB across the potential barrier created by band bending. The Fermi level in silicon is indicated by a dashed red line.} 
\label{CartoonMechanisms}
\end{figure}

\section{Non-equilibrium current through a DB state}\label{NE_current}
In the above discussion we assert that a net negative charge on the DB during imaging can only be explained by a non-equilibrium phenomenon, and is in fact impossible if the DB was in electrochemical equilibrium with the silicon crystal (due to the DB level being greatly elevated from the bulk Fermi level). In our system, we have a situation where the current flows from one reservoir (metal) at the chemical potential $\mu _{\rm tip} = \mu_1$ to another (Si crystal) at a chemical potential $\mu _{\rm Si}=\mu_2$ via a discrete level (DB) whose energy is somewhere between $\mu_{\rm tip}$ and $\mu_{\rm Si}$. An electron can be transferred into or out of the localized DB level at two different rates, say $r_{\rm tip} = r_1$ and $r_{\rm Si}= r_2$, which in general include all elastic and inelastic charge transfer mechanisms. Overall, a steady state of current flow is achieved when $r_{\rm tip}= r_{\rm Si}$ in which the occupation of the DB level depends on the DB energy and the escape rates. 

For a DB state in equilibrium with reservoir 1 only (or 2 only), its occupation equals the Fermi-Dirac distribution, $f_{\rm DB}^{(1)} =f_{\rm FD} (E_{\rm DB} ,\mu _{1} )=[1+\exp ((E_{\rm DB} -\mu _{1} )/kT)]^{-1} $ (or $f_{\rm DB}^{(2)} =f_{\rm FD} (E_{\rm DB} ,\mu _{2} )=[1+\exp ((E_{\rm DB} -\mu _{2} )/kT)]^{-1} $, respectively). 
However, since the DB level is in contact with both reservoirs, the DB will achieve a steady state of filling, with an occupation $f^*_{\rm DB}$ somewhere in between (the superscript * stands here for non-equilibrium occupation function). Then, the current through one contact equals the current through the other, $I_1=I_2$, which is the \textit{condition for a steady state occupation} of a DB. In order to calculate the steady-state current and DB occupation, we must first determine the flow rates between the tip and DB on one hand, and between sample and DB on the other. The possible mechanisms of charging and discharging a DB in the empty state STM imaging mode are shown in \fig{CartoonMechanisms}. 

Note that, although in principle the DB$^+$/DB$^0$ transition should also contribute to the transfer of electrons from the tip to the silicon crystal, its contribution can in practice be neglected for subtle reasons derived from the fact that the DB$^+$ level is resonant with the VB band (and therefore the DB$^+$ state has a negligible lifetime). As a consequence: (i) tunneling of electrons from the STM tip to the DB$^0$ level almost never happens, because it implies a preexisting DB$^+$ state; (ii) thermal emission of an electron from the DB$^0$ level involves an additional thermal barrier of 0.45 eV (or 18$k_{\rm B}T$), compared to emission from the DB$^-$ level, and it can be completely neglected. Therefore, the function $f^*_{\rm DB}$ signifies the statistical occupation of the $DB^-$ level alone, and varies between 0 (for a DB with a time-average of zero charge) and 1 (for a completely filled DB$^-$, on average). In the sections below we discuss each charge transfer mechanism shown in \fig{CartoonMechanisms} in detail.

\subsection{Rate of electron transfer between STM tip and an isolated DB}
During a STM experiment, a possible mechanism for electron transfer from the STM tip to a DB (and vice versa) is by elastic tunneling across the vacuum gap. Another possible mechanism is a two-step capture process in which an electron from the tip first tunnels into the Si crystal in the vicinity of the DB and then immediately gets captured by the DB before it has the chance to become an extended-state electron. Henceforth we call this mechanism ``STM vicinity capture of tunneling electrons by a DB''.

\begin{enumerate}
\item  The rate of elastic transfer between the STM tip and an isolated DB can be estimated (see Appendix~\ref{App1} for more details) 
\begin{equation}	
I_{\rm tip-DB}(d) = \frac{4\pi e}{\hbar} \frac{1}{\Omega_{\rm DB}} \rho_{\rm tip}(E_{\rm DB}) [f_{\rm tip}(E_{\rm DB}) - f^{*}_{\rm DB}] |\tilde{M}_{\rm tip-DB}(E_{\rm DB},d)|^2,
\label{ItipDB}
\end{equation}  
where we denote the Fermi-Dirac distribution simply by $f$ with a subscript indicating the respective electrode (``tip'' for tip, or ``Si'' for silicon crystal), $\Omega_{\rm DB}$ is the volume of the DB (defined as the region in space where the DB charge density is greater than 1\% of its maximum), and $|\tilde{M}(E_{\rm DB},d)|^2$ is a properly normalized \textit{transfer matrix element} for separation $d$ (between DB and tip centers) defined as
\begin{equation}
  |\tilde{M}_{\rm tip-DB}(E_{\rm DB}, d)|^2 = \left(\frac{\hbar ^{2} }{2m} \right)^{2} \left|\int _{S_{0} }\mathbf{dS}\cdot  [\varphi _{\rm tip} \nabla \varphi _{\rm DB}^{*} -\varphi _{\rm DB}^{*} \nabla \varphi _{\rm tip} ]\right|^{2}.
\end{equation}
Note that in eq. \ref{ItipDB} above, we also replaced the Fermi-Dirac distribution function for the DB with the non-equilibrium occupation function. 
As explained in Appendix~\ref{App1}, the functions $\varphi$ above represent dimensionless wavefunctions of the respective states under consideration defined as
\begin{equation}
	\varphi(\mathbf{r}) = \sqrt{\Omega} \psi(\mathbf{r}),
\end{equation}
where $\psi$ is the normalized wavefunction and $\Omega$ is the volume of the electrode or the localized state, respectively. For the purpose of estimating the tunneling currents, the wavefunction of the STM tip is taken in the conventional form of the Tersoff-Hamann approach
\begin{equation}
  \psi _{E,\mathbf{K}}^{\rm tip} (\mathbf{r})=c_{\rm tip} (R_{0} )\frac{\exp [-\zeta _{\rm tip}(E) |\mathbf{r-r_{\rm tip}}|]}{\zeta_{\rm tip}(E) |\mathbf{r-r_{\rm tip}}|}
  \label{psiTip} 
\end{equation}
for $|\mathbf{r-r_{\rm tip}}|>R_{0} $, where $\mathbf{r_{\rm tip}}$ is the position of the tip center, $R_0$ is a characteristic tip radius, $c_{\rm tip} (R_{0} )=c_{\rm 0t} \zeta _{\rm tip} R_{0} \exp (\zeta _{\rm tip} R_{0} )/\sqrt{\Omega _{\rm tip} } $, $c_{\rm 0t}$ is a dimensionless constant of the order of unity, $\zeta_{\rm tip}(E) =\sqrt{2m(E_{\rm vac} -E)} /\hbar $, and $E_{\rm vac}$ the vacuum level.

Generally speaking, one would expect sequential tunneling events from the STM tip, first into the DB$^0$ level and then into DB$^-$ level. However, for an electron to tunnel into the DB$^0$ level, the DB has to be in a pre-existing DB$^+$ state, and the probability of having that state is negligible because of it being resonant with VB levels. Therefore tunneling into the DB$^0$ level can be neglected.

\item The STM vicinity capture of tunneling electrons by a DB involves tunneling from the STM tip into the Si crystal for a range of electron energies and momentum orientations and subsequent capture of those hot electrons by a neutral DB. The corresponding current into the DB is denoted by $C^{\rm STM}_{\rm n}$ and the details of its calculation are described in Appendix~\ref{cSTMappendix}.
\end{enumerate}

During an STM experiment, in addition to the tunneling current from the tip to the DB, other transfer mechanisms contribute to the overall balance of charge on a DB orbital. They are discussed in detail below below.

\subsection{ Rate of electron transfer between an isolated DB and the silicon crystal}
Once an incoming electron becomes localized at a DB state, there are three main mechanisms of transfer into the silicon crystal:
\begin{enumerate}
\item  The localized DB electron escapes by emission into the conduction band via thermal excitation. This thermal escape rate only depends on the barrier height and temperature; the escape rate is given by 
\begin{equation}
	e_{\rm n} =\sigma _{\rm n} v_{\rm n} N_{c} \exp [-(E_{\rm CBM} -E_{\rm DB} )/k_{\rm B} T],
	\label{e_n}
\end{equation}
where the prefactor is an attempt frequency and $N_c$ is the effective density of states at the bottom of the conduction band
\begin{equation}
	N_c=\frac{1}{\sqrt{2}} \left( \frac{m_{\rm n,eff} k_{\rm B}T}{\pi \hbar^2}\right)^{3/2}
\end{equation}
with $m_{\rm n,eff}$ being the effective mass of electrons in Si. 

\item  The localized DB electron tunnels \textit{elastically} into the bulk CB level (at the classical turning point level where the CBM intersects the DB$^-$ level). Henceforth, CBM stands for CB minimum, and VBM stands for VB maximum. Tunneling from DB occurs across the space charge layer and the tunneling distance depends, via band banding, on the doping level of the sample, STM tip height and bias voltage.  The tunneling current can be estimated by using the formula (see Appendix\ref{App1})
\begin{equation}
	I_{\rm DB-Si} = \frac{4\pi e}{\hbar} \frac{1}{\Omega_{\rm DB}} g_{\rm Si}(E_{\rm DB}) [ f^{*}_{\rm DB}- f_{\rm Si}(E_{\rm DB})] |\tilde{M}_{\rm DB-Si}(E_{\rm DB})|^2,
\end{equation}
where $\tilde{M}_{\rm DB-Si}$ in this case also depends on the tunneling barrier for an electron with energy $E_{\rm DB}$ which is given by the band bending dependence on $z$. Upon tunneling into the bulk, the DB resumes a neutral state.

To complete the picture, for the purpose of calculating transfer coefficient between the tip and the silicon sample, we assume sample wavefunctions derived from the ``jellium model''\cite{leavens1988tunneling} with modifications to account for the corrugation of the surface
\begin{equation}
  \psi _{E,\mathbf{K}}^{\rm Si} (\mathbf{r})=- c_{\rm Si} \frac{k_{z} \exp \left\{-\sqrt{\zeta _{\rm Si}^{2} +K^{2} } [z-H(\mathbf{R})]\right\}}{ \sqrt{\zeta_{\rm Si}^{2} +K^{2} } -ik_{z} } \exp (i \mathbf{K} \cdot \mathbf{R}),
  \label{psiSam}
\end{equation}
where $E$ is the eigenenergy of the state; $\mathbf{K}= (k_x, k_y)$ is the surface parallel wavevector; $k_z$ is the surface normal wavevector (the last three being related via $E=\hbar ^{2} (K^{2} +k_{z}^{2} )/2m$); $\mathbf{r}=(x,y,z)$,  $\mathbf{R}=(x,y)$; $\Omega_{\rm Si}$ is the sample volume; $\zeta_{\rm Si}^{2} =2m(E_{\rm vac} -E)/\hbar ^{2} $; and $c_{\rm Si} = i c_{\rm 0s} / \sqrt{\Omega_{\rm Si} }$ with $c_{\rm 0s}$ a dimensionless constant of the order of unity. The corrugation function of the sample $H(\mathbf{R})$ is in general two-dimensional, but in our model we treat the $x$ and $y$ directions as independent. We assume the corrugation component along dimer rows (for which we calculate STM current traces) has the form $H(x)= A \cos^2(\pi x/L_x)$, with $A$ the corrugation amplitude and $L_x$ its period.

\item  A hole striking the surface is captured by the occupied DB level (defect mediated recombination). The associated recombination current is calculated in a customary approximation, by assuming that the rate of recombination at a DB site is proportional to concentration of holes, $p$, at the vicinity of the DB, the thermal velocity of those holes, $v_p$, and a capture cross section, of a hole by a negatively charged DB, $\sigma_{\rm p} $: 
\begin{equation}
	r_{\rm rec} = \sigma _{\rm p} p v_p 
\end{equation}
Their numerical values are $v_p = 1.87\times10^7$ cm/s, $\sigma _{\rm p} =1.1$~\AA$^2$, while $p$ depends on the band bending at the surface and can be anywhere from 70 cm$^{-3}$ in the bulk of highly doped n-type Si (resistivity 0.01~$\Omega$-cm) to $10^{17}$ cm$^{-3}$ in an inversion layer. The recombination current at the DB is given by
\begin{equation}
  I_{\rm rec} = C_{\rm p} = e \sigma _{\rm p} p v_p f^*_{\rm DB}
\end{equation}
\end{enumerate}
An additional note of the capture cross section: defects in semiconductors, both in bulk and at surface/interface, have been known to act as very effective recombination centers. In fact, carrier recombination in a semiconductor is actually dominated by trapping processes at defects, such as DB states. An important fact is that defects with energies close to band edges (as opposed to the middle of the band gap) are more efficient at trapping carriers and causing recombination \cite{moison1987surface}. Also, the above rough values of recombination rates rely on the semiclassical estimates for the hole concentration at the surface and include no charge quantization effects due to band bending. This quantum effect can significantly alter the carrier concentration at the surface and therefore can be important for the recombination rates.

Generally speaking, the capture cross section has a thermally activated behavior, 
\begin{equation}
  \sigma _{\rm p} =\sigma _{0p} \exp (E_{\rm a} / k_{\rm B}T)
\end{equation}
where  $\sigma _{0p} $ is a constant and $E_{\rm a}$ is an activation energy. For our purpose, we note that the capture cross section of a DB is more or less independent of temperature between 100-400 K \cite{lanno2}, for both electron and hole capture. 
It follows, that in the absence of tunneling, the flow of electrons in/out of a DB is given by
\begin{equation}
  R_{\rm n} = C_{\rm n} - E_{\rm n}= e\sigma _{\rm n} v_{\rm n} n(1- f^*_{\rm DB} ) - e e_{\rm n} f^*_{\rm DB} 
\end{equation}
with capture competing with emission, and the unknown occupation, $f^*_{\rm DB}$ dictated by the non-equilibrium conditions.

\section{Solving for the non-equilibrium occupation of a DB}
Combining all the above mechanisms we obtain the total current flow between DB and bulk
\begin{equation}
  I_{\rm DB}^{\rm in} =I_{\rm DB}^{\rm out},
\end{equation}
yields
\begin{equation}
  I_{\rm tip-DB} + C^{\rm STM}_{\rm n} + C_{\rm n} = I_{\rm DB-Si} + E_{\rm n} + C_{\rm p},
\end{equation}
where the additional current components, in Ampere, are given by
\begin{eqnarray}
  C_{\rm n} &=& e \sigma _{\rm n} v_{\rm n} n (1- f^*_{\rm DB} ), \\ 
  E_{\rm n} &=& e e_{\rm n} f^*_{\rm DB}, \\
  C^{\rm STM}_{\rm n} &=& e \sigma_{\rm n} I_{\rm tip-Si}^{\rm DB^0} (1- f^*_{\rm DB} ) \int_{\Omega_{\rm DB}} d\mathbf{r} |\psi_{\rm DB}(\mathbf{r})|^2   \frac{D(\theta(\mathbf{r}))}{2\pi r^2}, \\
  C_{\rm p} &=& e \sigma _{\rm p} v_{\rm p} p f^*_{\rm DB},
\end{eqnarray} 
where $v_p$ is the thermal velocity of holes and $N_v$ is the effective density of states at the top of the valence band.
\begin{equation}
	N_v=\frac{1}{\sqrt{2}} \left( \frac{m_{\rm lp,eff} k_{\rm B}T}{\pi \hbar^2}\right)^{3/2}
	+ \frac{1}{\sqrt{2}} \left( \frac{m_{\rm hp,eff} k_{\rm B}T}{\pi \hbar^2}\right)^{3/2}
  \label{N_v}
\end{equation}
with $m_{\rm lp,eff}$ and $m_{\rm hp,eff}$ being the effective masses of light and heavy holes in Si.

The solution is
\begin{equation}
  f^*_{\rm{DB}} =\frac{i_{\rm tip} f_{\rm tip} (E_{\rm DB} ) + i_{\rm Si} f_{\rm Si} (E_{\rm DB} ) + c_{\rm n} + c^{\rm STM}_n  } {i_{\rm tip} +i_{\rm Si} + c_{\rm p}  +e_{\rm n} +c_{\rm n} + c^{\rm STM}_n } ,
\end{equation}
where the charge transfer rates, in s$^{-1}$, are given by
\begin{eqnarray}
c_{\rm n} &=& \sigma _{\rm n} v_{\rm n} n, \\ 
c^{\rm STM}_n &=& \sigma_{\rm n} \frac{I_{\rm tip-Si}^{\rm DB^0}}{e} \int_{\Omega_{\rm DB}} d\mathbf{r} |\psi_{\rm DB}(\mathbf{r})|^2   \frac{D(\theta(\mathbf{r}))}{2\pi r^2},  \\
c_{\rm p} &=& \sigma _{\rm p} v_{\rm p} p, \\ 
i_{\rm tip} &=& \frac{4\pi}{\hbar} \frac{1}{\Omega_{\rm DB}} g_{\rm tip}(E_{\rm DB})  |\tilde{M}_{\rm tip-DB}(E_{\rm DB})|^2 , \\ 
i_{\rm Si} &=& \frac{4\pi}{\hbar} \frac{1}{\Omega_{\rm DB}} g_{\rm Si}(E_{\rm DB}) |\tilde{M}_{\rm DB-Si}(E_{\rm DB})|^2,
\end{eqnarray}  
and where $D(\theta(\mathbf{r}))$ in $c^{\rm STM}_n$ is the normalized tunneling current density in the direction $\theta$ from the STM tip to the sample, which is derived in Appendix C.

The total time-average  STM current is then given by
\begin{equation}
  I_{\rm STM} =I_{\rm bulk}^{\rm DB0} (1- f^*_{\rm DB}) +I_{\rm bulk}^{\rm DB-} f^*_{\rm DB} +I_{\rm DB}^{\rm out},
  \label{totalI}
\end{equation} 
where $I_{\rm bulk}^{\rm DB0}$ and $I_{\rm bulk}^{\rm DB-}$ are the tip to sample currents in the presence of a neutral and negative DB, respectively.

From \textit{ab initio} calculations on isolated DBs at the surface of model silicon clusters\cite{haider1}, we know that a negatively charged DB renders its host Si atom to be slightly elevated from the plane of the surface, about 0.3~\AA, which can also have an effect on the tunneling coefficient during STM imaging. This effect is also captured in our model by accordingly modifying the sample wavefunction, and plays a role in the appearance of the DB. 

Another possible effect during DB imaging is a Stark shift of the DB level due to the tip-induced field in which the DB is found. This shift would be most pronounced when the field is stronger, i.e. when the tip apex is immediately above the DB, for a given sample bias. In principle, calculating the magnitude of this shift is possible within the frame of our current theory by solving the Schroedinger equation in the presence of the external field, e.g. by a perturbative approach or finite element method. This would produce another parameter in our model, namely the magnitude of the Stark shift as a function of the tip-DB distance and tip height. However, for the sake of simplicity, we decide not to include this additional parameter in our model, but rather to comment on its possible consequences on the model results.

Note that if the ``equilibrium would-be'' populations are equal the net current through the DB is zero (this only happens if the chemical potentials on the two sides are equal). Also note from above formulas that, if one of the flow rates is much smaller than the other, then the net current will be reduced to just that rate. As another general remark, an implicit assumption above is that the total STM current (through DB and bulk states) is small enough that it does not significantly change the Fermi level of the reservoirs in the vicinity of the two contacts.

\section{Results of the imaging model}
In this section we present the numerical results of the theory presented above for the imaging of DBs. We illustrate our model with experimental and theoretical results on a highly doped n-type silicon sample, with a H-terminated 2$\times$1 (001) surface reconstruction. We only focus on the unoccupied-state imaging mode of STM, where most of our experimental data was obtained. In this case, upward band bending occurs at the surface and an excess concentration of holes is present there. The STM tip injects electrons into the sample, and into isolated DBs as well. Localized electrons at DBs discharge into the sample by the different channels described above, including recombination with the excess holes at the surface.
Other sample doping type and levels are also amenable to this model, keeping in mind that the system parameters will be different and require making appropriate adjustments: Fermi level, band bending, local electron and hole concentrations at the surface, etc. For example, in the occupied-state imaging, electrons are extracted from the sample into the STM tip, and downward band bending occurs at the surface, which brings about an excess of electrons, as opposed to an excess of holes. This has consequences on the filling of isolated DBs, prior to their discharge to the STM tip.
\begin{figure}[tbp]
\centerline{
\includegraphics[scale= 1.0, width= 0.88\linewidth]{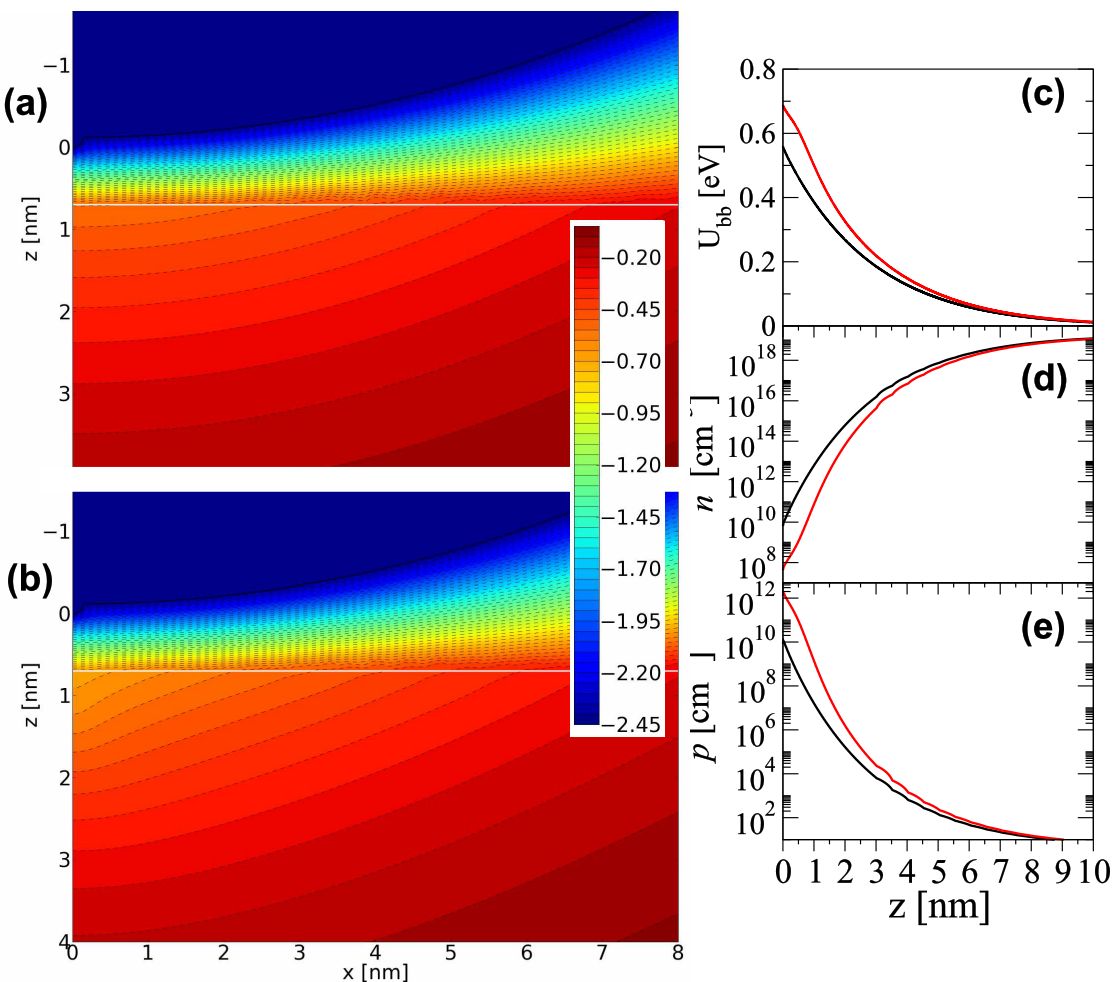}
}
\caption{
(a) Two-dimensional section of the electrostatic potential landscape caused by a biased STM probe in the vicinity of a silicon surface for unoccupied state imaging and for axial symmetry (tip central axis is at $x=0$). The potential penetrates the surface and vanishes deep in the crystal. The Si crystal is n-type with a resistivity of 3.5 m$\Omega$-cm, voltage bias on the STM tip is -2V, and the tip height was assumed to be 0.7nm (tip apex is at $z=0$). The tip contour is drawn as a black line and the surface as a white line.
(b) The electrostatic potential for the same biasing conditions as in (a) and in the presence of a negative DB directly under the tip apex. The color bar indicates the values of the electrostatic potential in V, and is identical for both plots for easy comparison.
(c,d,e) Subsurface band bending and carrier concentration along the central axis of the tip plotted as a function of sample depth, $z$, for biased tip in the presence of a DB$^0$ (black curves), and in the presence of a DB$^-$ (red curves) placed directly under tip apex.
} 
\label{fig2a}
\end{figure}

Finite element analysis was used for calculating tip-induced band-bending as the self-consistent solution of Poisson and Schroedinger equations~\cite{upcomingFEM}. This solution is reliable for the regime in which quantum confinement effects (quantization of electronic levels at surface) are negligible, but also for the case of strong inversion or accumulation, when quantum effects are important\cite{feenstra2003electrostatic}. In this latter case, many surface quantized levels are occupied, which makes the semiclassical approximation for the charge distribution valid. The semiclassical calculation fails to be reliable when the inversion or accumulation is moderate, that is, when surface quantized states exist but only a few quantum states are occupied. Some details of the FEM calculation are given in Appendix \ref{FEMappendix} and the complete details of the whole calculation procedure are described elsewhere~\cite{upcomingFEM}. 

In Figure~\ref{fig2a} we show an example of calculated potential landscape (by FEM) for chosen sample bias and tip height as (a,b) a two-dimensional section of the axially symmetric system, and (c) a function of the depth inside the silicon surface. The effect of a negative DB under the tip on the band bending was also calculated and shown here. The greater differences between the two cases can be seen in the silicon region immediately under the tip apex, where the DB is located. Simultaneously, we calculate the carrier concentrations in the sample induced by the tip alone and also by the tip in the presence of a DB$^-$ and plot the results in panels (d,e).

Unlike the case of a highly doped p-type Si sample, where the main discharging rate of a DB$^-$ is that of recombination with holes~\cite{berthe2008probing}, in our case, both types of carrier concentration are small at the surface, where the sample is rendered almost intrinsic by the tip-induced potential. As a consequence, recombination is no longer the dominant discharging rate and it competes with the other rates described above.

\begin{figure}[tbf]
\centerline{
\includegraphics[scale= 1.0, width= 0.65\linewidth]{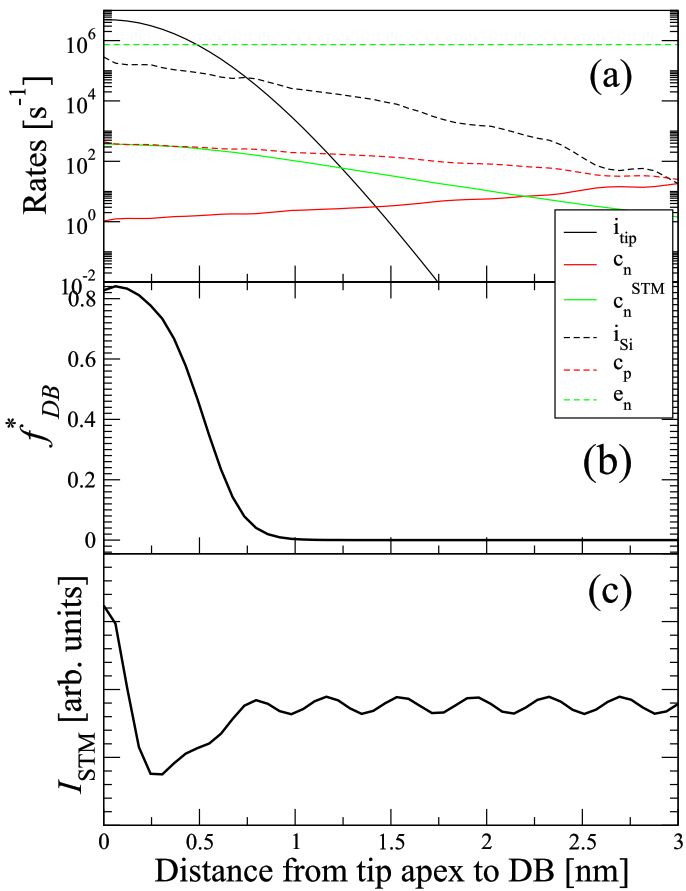}
}
\caption{Dependence of relevant physical quantities on the surface-parallel distance between the STM tip apex and the center of an isolated DB. Voltage bias on the sample is +2V and the tip height is 7~\AA~ from the surface. (a) The dependence of various elastic and inelastic rates of charge transfer through a DB in the presence of an STM imaging tip. (b) Non-equilibrium occupation function of the DB$^-$ state versus distance. (c) Total STM current dependence on the same distance. This situation corresponds to the  experimental Figure \ref{experimImages1}(a), where a halo is formed.} 
\label{fig7a}
\end{figure}

\begin{figure}[tbf]
\centerline{
\includegraphics[scale= 0.2, width= 0.65\linewidth]{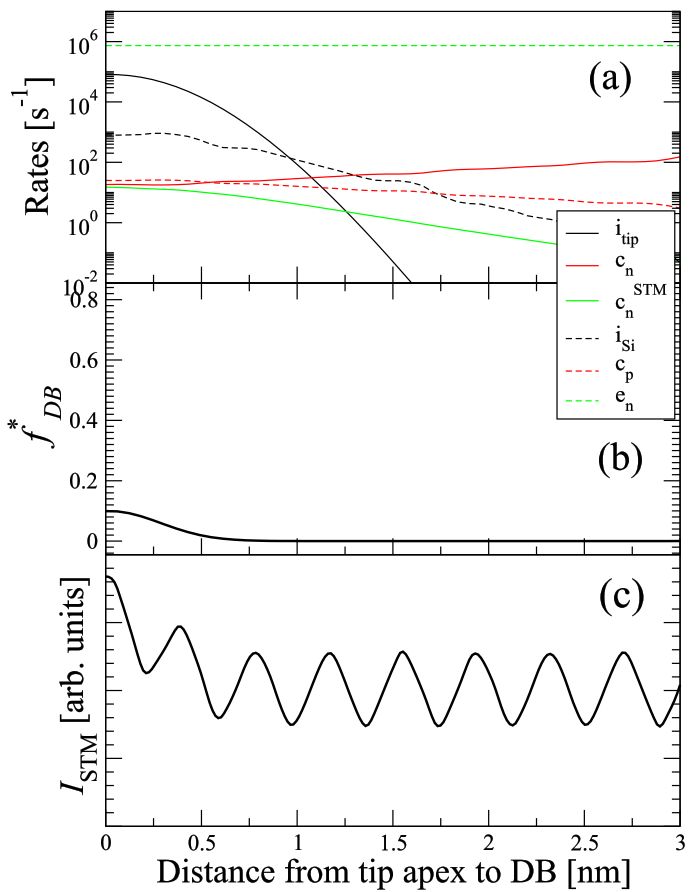}
}
\caption{The same quantities are plotted as in figure \ref{fig7a}, but for a different tip height, 8.5~\AA, which corresponds to the case in the experimental Figure \ref{experimImages1}(c), where halo is not formed.} 
\label{fig7b}
\end{figure}

The calculated band bending and carrier densities are input to the mathematical formalism described above in order to compare our theory to our experimental observations. As suggested above, the key point is to determine the charging state of the DB. 
The main results needed to understand the non-equilibrium charging effect are shown in Figures~\ref{fig7a} and \ref{fig7b}. Part (a) illustrates the competitions between electron transfer rates in and out of an isolated DB in the unoccupied-state imaging mode. As the STM tip moves in from far away toward the DB, the electron injection rate from tip to DB (solid black curve) increases approximately as an exponential function, up to a critical distance $r_c$ of about 5~\AA. At that point, this injection rate becomes greater than the dominant rate of transfer out of DB, which in this case is provided by the thermal emission of electrons into the conduction band. The second greatest discharge rate is by tunneling into resonant CB levels, and the third is by capture and recombination with a itinerant hole, $c_{\rm p}$, which is considerably smaller. 

For smaller distances than $r_c$, the discharge rate becomes the limiting rate for current  through the DB, and therefore it dictates the charging state of the DB, $f^*_{\rm{DB}}$, plotted in part (b). In turn this charging state determines the total band bending (tip-induced plus DB-induced) and also the tip-sample tunneling coefficient  in the vicinity of the DB and it therefore affects the total current measured in the STM scan, see part (c). Over a relatively short distance, in the vicinity of $r_c$, the time-average DB occupation switches from zero to one, because the electrons localized there cannot escape at the same rate at which they are injected. With the addition of a localized negative charge at the DB center, the electrostatic landscape changes in its vicinity, i.e. upward band bending increases, and the STM current is affected, giving rise to a dark halo with the outer edge around $r_c$. this corresponds to the situation depicted in the experimental image in Figure \ref{experimImages1}(a). The bright spot at the DB center is caused by the slight elevation of the host silicon atom at the location as mentioned above, which although small around 0.3~\AA, has a significant effect on the tunneling coefficient due to the exponential sensitivity of the latter with tip-sample separation.

A halo does not always appear around an isolated DB, and in our case this happens at lower injection current, as in the situation shown in Figure~\ref{experimImages1}(c), which is consistent with a greater tip height. To simulate the situation in our model, who performed the same calculations for a tip height of 8.5~\AA~ and the results are shown in Figure \ref{fig7b}. We can see a decrease in this case of the STM injection rate into the DB, which becomes lower than the thermal emission rate as seen in (a). Therefore the DB is mostly neutral in this situation, as seen in (b). No halo is formed here because additional band bending by the DB does not occur. 

\begin{figure}[tbf]
\centerline{
\includegraphics[scale= 0.2, width= 0.8\linewidth]{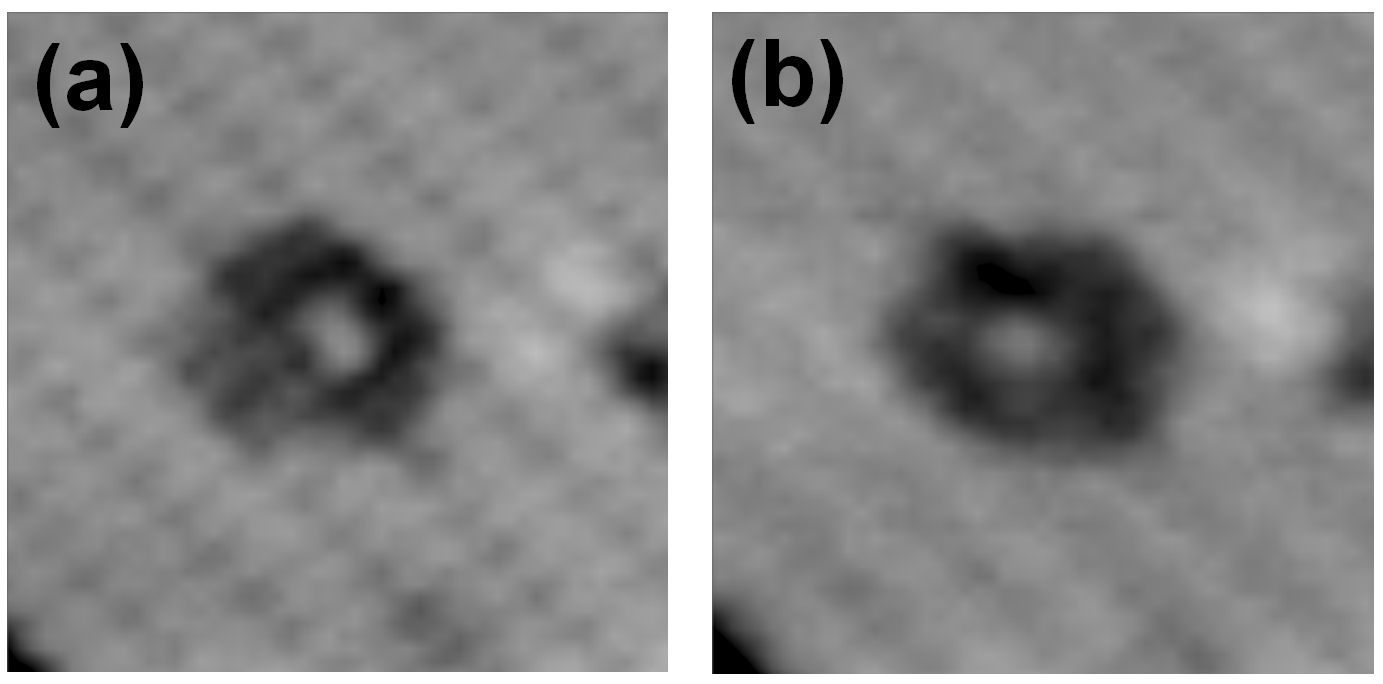}
}
\caption{Dependence of the shape of a DB halo on the particular shape of the STM tip used for imaging. The same DB was imaged in (a) and (b), at the same imaging setpoint (sample voltage 2~V, imaging current 0.2~nA), but with a tip change between (a) and (b).} 
\label{SameDBdifferentTips1}
\end{figure}

Over the course of many experiments, we also observed other related imaging trends of DBs: (i) there is a tip dependence in the exact shape of a DB halo, see \fig{SameDBdifferentTips1}; (ii) halo-like features around groups of two or more tunnel-coupled DBs. The dependence in (i) can be qualitatively described by the fact that different STM tips have atomic size protrusions at different locations in the vicinity of the apex. During scanning, as the tip approaches a DB, there are different onset points for complete charging ($f^*_{\rm DB} =1$) as the smallest distance between tip and DB  (and the corresponding tunneling rate) depends on the location of the protrusions for each tip. 

For (ii), the same imaging mechanism is at work for groups of coupled DBs, except the localized charge is shared among more than one DB, and the charge per atom is less~\cite{haider1}. Consequently, the DB-induced band bending tends to be lower. Also, due to the shape of the STM tip, the tip-induced band bending amounts to different values at different DBs, and therefore their levels are pulled upward with respect to the bulk VBM by different amounts. The localized charge tends to reside at the state whose energy level is lower (farthest from the tip).

Other possible effects were also taken into account in our analysis. The calculation of the charge carrier densities at the surface accounting for \textit{charge quantization} in the surface-perpendicular direction was also performed in a one-dimensional model ($z$-dependence only) by solving the Schroedinger equation for the quantum well formed by the band bending in the $z$-direction. However, for the situation depicted in Figure \ref{fig2a}, band bending is not strong enough to induce significant quantization of the space charge layer, and the semiclassical treatment is reliable. 

As we mentioned above, during a STM scan when the tip apex is directly above the DB, the electric field under the apex can be very strong (around 2 V/nm in the vacuum gap) and this can induce a Stark shift of the DB level and a deformation of the DB orbital. This shift is of the order of a few tens of meV (from our preliminary estimates) and is not included as a parameter in our model. However, this shift has possible consequences on the appearance of DB, especially in the region close to the DB center. An upward shift in $E_{\rm DB}$ would decrease the band bending barrier and thus increase the $i_{\rm Si}$ rate, and also the thermal excitation rate $e_{\rm n}$ (these rates are very sensitive to the exact DB level). This means that more current would pass through the DB state itself, contributing to a brighter peak at the DB center. We expect this effect to become significant only for images taken at lower tip heights (higher setpoint currents).

Band bending in a semiconductor is usually explained by appealing to semiclassical arguments. However, for the semiclassical theory to be accurate, one must involve the approximation that the perturbation in the crystal varies slowly on the spatial scale of the crystal lattice constant. In order to test this approximation for our case, here we also undertake a more general approach that can be used beyond the semiclassical approximation, which can also cast insight into the long range behavior of screening of a negative DB. For this purpose, the calculation of the electrostatic potential landscape was done by separating the whole problem into two components, considered to be approximately independent, according to the source of electrostatic perturbation: (i) the tip-induced band bending; and (ii) band bending by an isolated charged DB alone. Solving these two problems requires significantly different approaches caused by the different spatial scales of the problems. A perturbation approach~\cite{march2, march3, march6} was used for the problem of DB-induced band bending. This approach includes the semiclassical approximation and the random phase approximation (RPA) as particular limits, and allows us to analyze other quantum mechanical effects, such as Friedel oscillations.

\begin{figure}[tb]
\centerline{
\includegraphics[scale= 0.2, width= 0.65\linewidth]{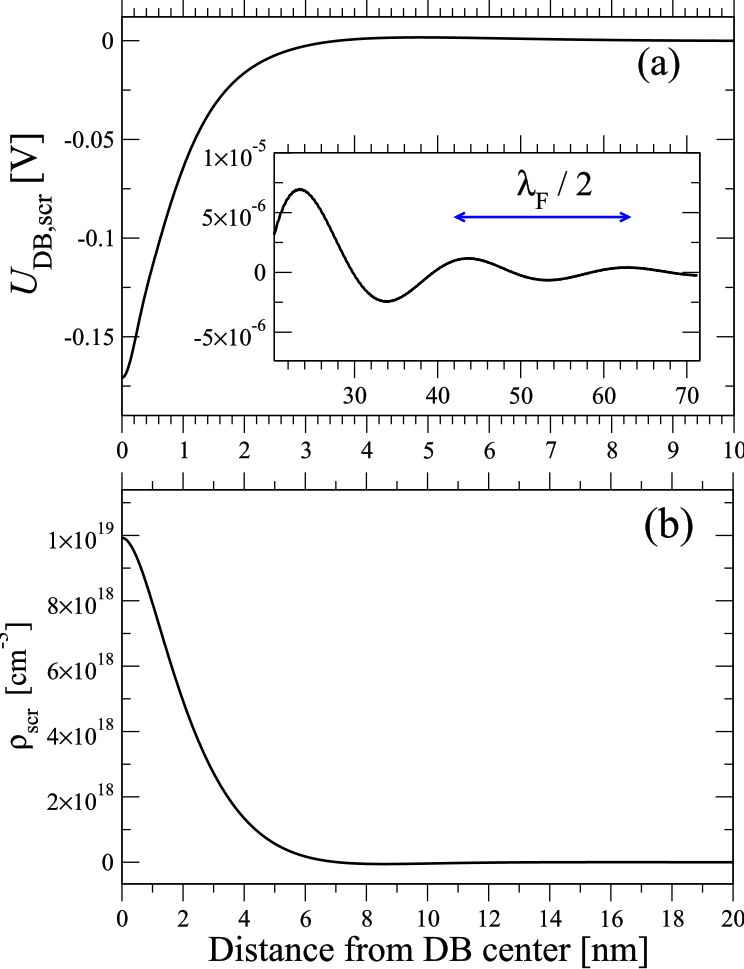}
}
 \caption{(a) The screened electrostatic potential solution for a negatively charged single DB as a function of distance from the DB center calculated using the perturbation approach ~\cite{march2, march3, march6}. Inset: the long range behavior of the potential displaying thermally damped Friedel oscillations, albeit with much lower amplitudes than the potential at close range (the horizontal axis is the distance in nm for all plots). (b) Charge density of the mobile carriers (holes) brought about to screen the DB charge.}
\label{fig4a}
\end{figure}
In order to make the computational task feasible for long spatial range (100 nm), the angular dependence of the DB orbital was integrated out in this case (with only radial dependence remaining). The DB is considered to be placed in a background charge density determined by the tip-induced band bending evaluated by FEM calculations. In Figure~\ref{fig4a}, we show the solution of the (separate) electrostatic problem of screening of a DB by the mobile carriers in the semiconductor. The maximum band bending due to the DB$^-$, is around 0.17 eV, which is slightly larger than the all-semiclassical results presented above. Thermally damped Friedel oscillations are indeed present in our results (inset in (a)), but they cannot be associated with the appearance of dangling bonds, because: (i) they are extremely small in magnitude compared to the other effects discussed in our model (by more than four orders of magnitude); and (ii) they are manifest at greater distances ($>$ 20 nm) than the length scales associated with imaging halos.

Over the course of many experiments, we noticed a significant effect of the sample preparation methods on the way DBs image. In particular, there is a strong correlation between the value of the temperature used for flash-heating the sample and the formation of (or lack of) dark halos around DBs. This effect is caused by the fact that at high flash heating temperatures ($>$ 1250$^o$~C), there is a substantial migration of dopant atoms away from the surface region. This dopant depletion happens up to a depth ranging from hundreds of nm to a few microns and alters the local Fermi level at the surface. In turn, this has consequences on the charge transfer rates discussed above and ultimately on the charging state of the DB under imaging conditions. These facts also suggest  possible control mechanisms of the charging state of a DB, by adjusting the local Fermi level at the surface, e.g. by applying an external potential via a local electrode. This was also illustrated and discussed in a recent study on groups of closely coupled DBs on hydrogen terminated silicon surfaces~\cite{pitters2011tunnel}. 

\section{Summary and Outlook}
In this paper we propose a mechanism for understanding the STM imaging of isolated DBs on silicon surface, with quantitative focus on the n-type surfaces and unoccupied-state imaging mode. A quantitative analysis of the electrostatics of our system shows that the tip-induced band bending plays an important role in determining the charging state of the isolated DB by effectively pulling the DB state out of equilibrium with its host crystal. Under these conditions, the equilibrium Fermi-Dirac occupation of a DB is replaced by an occupation function dictated by the steady-state flow of current through the DB, which is very sensitive to the lateral DB-tip distance. Therefore the charging state of the DB can vary quickly upon lateral approach of the scanning tip, thus causing halo-like features with sharp edges around the DB. The sharpness of the edge is related to the slope of the variation of the DB occupation function $f^*_{\rm DB}$ with the lateral distance, see Figure~\ref{fig7a}. 

We also found that the all-semiclassical approach to band bending by an isolated DB$^-$ compares reasonably well, in terms of excess band bending by a DB$^-$, to more general approaches, specifically the RPA, and that Friedel oscillations do not account for imaging features present in this system. 

As the dominant DB discharging mechanism in our specific case study is the thermal emission of electrons into the silicon CB, lowering the system temperature should have a drastic effect on the DB image, and the halo should be increased. The similarity of our system to double barrier tunneling junction can also be exploited in certain regimes, e.g. by lowering the temperature and decreasing the thermal emission, thus rendering the DB-Si tunneling as the dominant discharging rate. 
These will be the subject of a further experimental investigation.
 
Our model calculates the total STM current in a time-average view of the system. If the time resolution of the STM was high enough, it should be possible to discern a ``telegraph signal'' of the current inside the halo region of an isolated silicon DB, as the band bending fluctuates with the charging/discharging of the DB. Our model does not reproduce the STM tunneling currents in a strictly quantitative fashion. However it does capture all the relevant mechanisms at play in image formation, and the main qualitative trends that were observed experimentally, with the variation of system parameters such as doping level, sample bias, and current setpoint. Because the  charging and discharging rates have exponential sensitivities to band bending values and distances between electrodes and localized states, their particular balance can be greatly affected by small changes in these parameters. Therefore imaging of DBs will be different for different semiconductor materials (Ge, GaAs), different surface reconstructions (e.g. Si(111)), for  samples doped at lower or higher levels, or p-type, and for different tips. The latter is the cause of daily variations in the appearance of DBs in the lab. However the general formalism presented here can be adapted to suit those similar systems.

\emph{This work is supported by iCORE, the National Research Council of Canada, NSERC, and CIfAR.}

\newpage

\begin{appendices}
\section{Formulas for calculating tunneling currents}\label{App1}
Throughout this study we use a generalized formula\cite{cutler} for tunneling currents between two electrodes (1 and 2) of volumes $\Omega_1$ and $\Omega_2$ in the presence of a externally applied voltage, $V_{\rm bias}$, equal to the difference between the chemical potentials on the two sides $\mu_1 -\mu_2$
\begin{eqnarray}
	I_{1\rightarrow2}
		    &=& \frac{4\pi e}{\hbar} \int dE [f(E, \mu_1) - f(E, \mu_2)] 
				g_1(E) g_2(E) |\tilde{M}(E)|^2.
	\label{generalI}
\end{eqnarray}
Here $g_1(E)$ and $g_2(E)$ are appropriately generalized local density of  states (LDOS) per unit energy \textit{and} unit volume (in units of J$^{-1}$m$^{-3}$) of the two electrodes;
 $|\tilde{M}(E)|^2$ is an appropriately defined (see below) transition matrix element between states in each electrode having energy $E$. We assume that our system fulfills the conditions in which the above formula is equivalent to the golden-rule expression for the tunneling current\cite{appelbaum} 
\begin{equation}
	I_{1\rightarrow2}= \frac{2\pi e}{\hbar} \sum_{m,n} [f(E_m, \mu_1) - f(E_n, \mu_2)] \delta(E_m - E_n) |M_{mn}|^2,
\end{equation}
where $E_m$ and $E_n$ are the energy levels of eigenstates $m$ and $n$ on each side with the (spin excluding) wavefunctions $\psi_{m}$ and $\psi_{\rm n}$ respectively. In a first order perturbation theory the transition matrix element was shown to be \cite{bardeen1961tunnelling}
\begin{equation}
 |M_{mn} |^{2} =\left(\frac{\hbar ^{2} }{2m} \right)^{2} \left|\int _{S_{0} }dS\cdot  [\psi_{m} \nabla \psi _{\rm n}^{*} -\psi _{\rm n}^{*} \nabla \psi _{m} ]\right|^{2} .
\end{equation}
where the surface integral is carried out over a median surface $S_0$.

If we define dimensionless wavefunctions $\varphi$, instead of normalized ones $\psi$, by
\begin{equation}
 \varphi_{m} (\mathbf{r}) = \sqrt{\Omega_1} \psi_m (\mathbf{r}),
\end{equation}
then we can further write
\begin{equation}
 |M_{mn} |^{2} =\frac{1}{\Omega _1 \Omega _2 }\left(\frac{\hbar ^{2} }{2m} \right)^{2}  \left|\int _{S_{0} }\mathbf{dS}\cdot  [\varphi _{m} \nabla \varphi _{\rm n}^{*} -\varphi _{\rm n}^{*} \nabla \varphi _{m} ]\right|^{2},
\end{equation}
and define a volume-independent transfer matrix element as
\begin{eqnarray}
 |\tilde{M}_{mn}|^2
     &=& \Omega_1 \Omega_2 |M_{mn}|^2
     =\left(\frac{\hbar ^{2} }{2m} \right)^{2}  \left|\int _{S_{0} }\mathbf{dS}\cdot  [\varphi _{m} \nabla \varphi _{\rm n}^{*} -\varphi _{\rm n}^{*} \nabla \varphi _{m} ]\right|^{2},
\label{Mmntilde}
\end{eqnarray}
which appears in eq. \ref{generalI} and fulfills the correct dimensionality of the integral.

We apply the above formalism for calculating the tunneling current for the three tunneling cases in our system: (i) between the STM tip and the silicon sample; (ii) between  the STM tip and an isolated DB; and (iii) between an isolated DB and the silicon sample. Below we adapt the general formula above to each of these cases in order to  derive the simpler forms given in the text.

For the first case, we assume that the tip wavefunctions are given as in the Tersoff-Hamann approach by eq. \ref{psiTip} and its LDOS is constant in energy. Also,  we assume that the sample LDOS is constant in space and has the customary square root dependence on energy for the bands in semiconductors
\begin{equation}
	g_{\rm Si} (E) = \frac{8\pi \sqrt{2}}{\hbar^3} m_{\rm n,eff}^{3/2} \sqrt{E- E_{\rm CBM}}
	\label{g_CB}
\end{equation}
for the conduction band levels (above CBM) and
\begin{equation}
	g_{\rm Si} (E) = \frac{8\pi \sqrt{2}}{\hbar^3} m_{\rm p,eff}^{3/2} \sqrt{E_{\rm VBM} - E}
	\label{g_CB}
\end{equation}
for the valence band levels (below VBM), with $g_{\rm Si} (E)$ being zero in the band gap ($m_{\rm n,eff}$ and $m_{\rm p,eff}$ are the effective masses for electron and hole in silicon).

For the tunneling between a DB and an electrode (tip or sample), we take advantage of the fact that the LDOS of the DB is a $\delta$-function in energy, namely $g_{\rm DB}(E)= \delta(E-E_{\rm DB})/ \Omega_{\rm DB}$, and therefore  the integral over energy in eq.\ref{generalI} is reduced to a single-energy term for $E_{\rm DB}$. This yields a simplified formula
\begin{equation}
	I_{\rm tip-DB} = \frac{4\pi e}{\hbar} \frac{1}{\Omega_{\rm DB}} \rho_{\rm tip}(E_{\rm DB}) [f_{\rm tip}(E_{\rm DB}) - f^{*}_{\rm DB}] |\tilde{M}_{mn}(E_{\rm DB})|^2,
\end{equation}
where $\Omega_{\rm DB}$  is the volume of the DB, defined in our case as the region in space where the DB charge density is greater than 1\% of its maximum. A completely similar equation can be written for the tunneling between DB and Si just by replacing the subscript ``tip'' above with ``Si''. In this latter case, $\tilde{M}_{mn}$ depends on the tunneling barrier between the DB and CB states lying at the same energy, which is given by the form of the band bending in the Si region surrounding the DB. Using the semiclassical approximation, we assume the tails of the wavefunctions for states in the conduction band in the barrier region scale as $\varphi_{\rm CB} (z)= \exp(-\zeta_{\rm CB} z)$, with $\zeta_{\rm CB}=\sqrt{2m\bar{U}_{\rm bb}}/\hbar$ and $\bar{U}_{\rm bb}$ the average tunneling barrier height between DB and CBM. 

For all cases described above, the tunneling matrix elements were calculated numerically by assuming the corresponding wavefunctions given in the text.

\section{Calculating band bending in the vicinity of a DB during STM imaging}\label{FEMappendix}
Tip-induced band bending was calculated within the semiclassical approximation using finite element methods (FEM). We assume that the STM tip has an axially symmetric shape ending in a hemispherical surface of radius $R_{\rm tip}=20~{\rm nm}$. On top of this surface we assume the existence of an atomic-sized protrusion of radius $R_0$, which coincides with the spherical tip radius assumed in the Tersoff-Hamann model for the tip wavefunction. 

For an STM system, the total potential drop between the tip and the sample 
\begin{equation}
  (V_{\rm tip} -V_{\rm sam} )=V_{\rm bias} +\Phi _{\rm tip} -\chi -(E_{\rm CBM} -E_{\rm F} )
\end{equation}
where $\Phi _{\rm tip}$ and $\chi$ are the tip workfunction and electron affinity of Si. Another component of $U_{\rm BB}$ can arise in the vicinity of a charged DB from the field created by the localized electron.

The calculation involves solving a system of non-linear equations consisting of the Poisson and Schroedinger (PS) equations for the electrostatic potential $U$ and the charge densities $(n,p)$ of electrons and holes, respectively, in the semiconductor. Poisson equation reads
\begin{equation}
	\nabla^2 U (\mathbf{r})= - \frac{e}{\epsilon_{\rm Si}} [N_D^+ + p(\mathbf{r}) - N_A^- - n(\mathbf{r}) ],
	\label{Poisson_eq}
\end{equation}
where $e$ is the elementary charge, $\epsilon_{\rm Si}$ is the electric permitivity of the semiconductor, $N_D^+$ the ionized donor concentration, and $N_A^-$ the ionized acceptor concentration. The same $U$ that satisfies the above equation, must also satisfy the Schroedinger equation
\begin{equation}
	- \frac{\hbar^2}{2m} \nabla^2 \psi (\mathbf{r}) + U(\mathbf{r}) \psi(\mathbf{r}) = E \psi(\mathbf{r})
\end{equation}
where, in our case, $\psi$ are extended states of the crystal belonging to the CB or VB bands, and their corresponding charge distribution determines $n$ and $p$.

In the semiclassical approximation, the extended states $\psi$ entering the above equation are affected by a local potential just by a \textit{rigid shift} in their energies equal to $-eU(\mathbf{r})$. As a consequence, the local charge densities are shown to be proportional to the (1/2)-Fermi-Dirac integrals, $F_{1/2}$
\begin{equation}
	\begin{array}{l}
		n(\mathbf{r}) = N_c F_{1/2}((E_{\rm F}- E_{\rm CBM}^{\rm bulk} + eU(\mathbf{r}))/ k_{\rm B}T)\\
		p(\mathbf{r}) = N_v F_{1/2}((E_{\rm VBM}^{\rm bulk} - eU(\mathbf{r}) - E_{\rm F})/ k_{\rm B}T)
	\end{array}
	\label{np}
\end{equation}
where $E_{\rm F}$ is the Fermi level (chemical potential) of the crystal. The above PS equations were solved self-consistently for the tip-induced band bending (no charge present on a DB) using an iterative FEM scheme in which the solution to eq. \ref{np} was input as a source term in the RHS of Poisson eq. \ref{Poisson_eq}. To ensure convergence of the iterations the Anderson mixing scheme was used. 

In order to calculate the distortions in the bands caused by the field of a DB$^-$, we performed a second FEM calculation in the presence of the charged DB. For practical reasons, we did not employ FEM on a three-dimensional geometry (for the case when the DB is laterally displaced from the central axis of the tip). Such a calculation is not practically feasible for the following reasons: (i) the size of the finite element domain must extend very deep into the semiconductor bulk in order to ensure that boundary condition there (zero electrostatic potential) does not introduce non-physical artifacts in the results; (ii) a very fine mesh is needed for the tip vicinity where the field is very strong in contrast to the bulk semiconductor where the field is relatively weak; (iii) many PS iterations are required to reach convergence, depending on the biasing conditions and doping level in the semiconductor.

We found a way to approximate the full solution of the PS system in the presence of a charged DB by taking advantage of the fact that the STM tip has a radius much greater than the size of the DB, and therefore it can be seen as locally flat in the vicinity of the DB. This implies that the \textit{potential due to the DB$^-$ alone}, $U_{\rm DB-}(\mathbf{r})$, will be approximately the same (as measured from the center of the DB) whether the DB is directly under the tip apex, or slightly displaced laterally, by a few nm. This allows us to extract $U_{\rm DB-}(\mathbf{r})$ as the difference between two different FEM solutions for two axially symmetric systems: (i) STM tip alone, and (ii) STM tip with a DB$^-$ directly under the apex. Then, for the case when the DB is not on the central axis of the tip but close to it, the solution is well approximated by adding the potential $U_{\rm DB-}(\mathbf{r-r_{\rm DB}})$ to the tip-induced potential.
More details of the finite element calculation will be published in an upcoming paper \cite{upcomingFEM}.

The calculation of the charge carrier densities at the surface \textit{including charge quantization} in the surface-perpendicular direction was also performed.  We numerically solved the Poisson and Schroedinger equations self-consistently for a planar metal-vacuum-semiconductor system in a one-dimensional case, assuming translational symmetry in $xy$-plane, in which case the Poisson and Schroedinger equations need only to be solved in the $z$-direction. This approximation is justified by the fact that the surface-parallel length-scale of the sample in which tip induced band bending is manifest is on the order of hundreds of nanometers, and much greater than the surface-normal length-scale over which the hole density is significant (tens of nanometers). In other words, the confinement of the holes in the potential well created by the biases STM probe in the surface-parallel direction is much weaker, than the confinement in the surface-normal direction. Charge quantization was also estimated using a 1D Poisson solution\cite{snider2009self} corresponding to the limit of a very large tip radius (or flat tip); those results were consistent with ours.

\section{STM vicinity capture of tunneling electrons by a DB}
\label{cSTMappendix}
The rate of capture for a deep-level state is proportional to the velocity of the electron being captured\cite{SHOCKLEY}, $ c_{\rm{n}} = \sigma_{\rm{n}} v_{\rm{n}} n   $. In equilibrium, at finite temperature, electrons sit at the bottom of the conduction band, and their energy distribution is well described by Maxwell-Boltzmann statistics, as long as $E_{\rm{CBM}}-\mu_{\rm{sam}}\gg k_{\rm{B}} T$. This allows the capture rate to be described by the familiar equation $ c_{\rm{n}} = \sigma_{\rm{n}} v_{\rm{th}} n   $, expressed in terms of the thermal velocity. The thermal velocity is the average group velocity for all electrons in the conduction band. Assuming equilibrium statistics and a parabolic conduction band minimum, the thermal velocity is given by $v_{\rm{th}}=\sqrt{8 k_{\rm{B}} T / \pi m^{*}}\approx 10^7$ cm/s.

However, in calculating the excess capture rate, $c_{\rm{n}}^{\rm{STM}}$, due to electrons injected from the tip into the conduction band, we no longer have recourse to Maxwell-Boltzmann statistics. We therefore use the group velocities directly. For simplicity, we retain the assumption of a parabolic band, so that $E_{\rm{\mathbf{k}}}=\hbar^2 k^2 /2 m$. In addition to simplifying calculations, this ensures that group velocities are parallel to their associated wavevectors. 

The injected electrons can tunnel into eigenstates in the conduction band with energies below $\mu_{\rm{tip}}$, however we expect that tunneling will occur most readily into states with high $k_{\rm{z}}$ (surface normal) and low $\mathbf{K}$ (surface-parallel) values. Thus their velocities are not evenly distributed across the range of polar angles, $\theta$. We solve for the angular distribution of tip-injected electrons by considering the matrix element, $\left| M_{mn}(\mathbf{k}) \right|^2$, for tunneling from tip states to sample states with wavevector, $\mathbf{k}$, described by equation \ref{Mmntilde}. The angular distribution of wavevectors, and hence also of velocities, of the injected electrons is then given by
\begin{equation}
D(\theta)=\mathcal{N} \int_{0}^{\sqrt{2m\left( \mu_{\rm{tip}}-E_{\rm{CBM}} \right)}/\hbar} k^2 \left| M_{mn}(\mathbf{k}) \right|^2 dk
\end{equation}
where
\begin{equation}
\mathcal{N} =\dfrac{2\pi }{ \int_0^{2\pi}  \int_{0}^{\pi/2}   \int_{0}^{\sqrt{2m\left( \mu_{\rm{tip}}-E_{\rm{CBM}} \right)}/\hbar} k^2 \left| M_{mn} \right|^2 sin\theta dk d\theta d\phi}.
\end{equation}

Ultrafast pump-probe reflectivity measurements of the Si(001) surface place the momentum relaxation time of free carriers at $32$ fs\cite{SABBAH}. Electrons traveling with speed $v_{\rm{th}}$ (much slower than the average velocity for injected electrons) will travel a distance of roughly $6$ nm in this time. We therefore make the approximation that injected electrons retain their initial group velocities over the distance scales relevant to the present problem. The local excess electron density at each point in the silicon sample, due to the injected current from the STM tip when the DB is neutral, $I_{\rm{tip-Si}}^{\rm{DB}^0}$, is then given by
\begin{equation}
n^{\rm{STM}}(\mathbf{r})=D(\theta)\frac{I_{\rm{tip-Si}}^{\rm{DB}^0}}{2\pi r^2v}
\end{equation}
where $\mathbf{r}=(r,\theta, \phi)$ being the distance vector from the STM tip apex, with the polar angle $\theta$ being measured from the tip axis pointing toward the silicon crystal.

Finally we account for the occupation of the DB with one electron, $1-f_{\rm{DB}}^{*}$, and write the capture current by the DB level as
\begin{equation}
C_{\rm{n}}^{\rm{STM}}=(1-f_{\rm{DB}}^{*})\int_{\Omega_{\rm{DB}}} d\mathbf{r} \left| \psi_{\rm{DB}}(\mathbf{r}) \right|^2 D(\theta) I_{\rm{tip-Si}}^{\rm{DB}^0} \frac{\sigma_{\rm{n}} }{2 \pi r^2},
\end{equation}
where we assume that the capture cross section for an infinitesimal volume is $\left| \psi_{\rm{DB}} \right|^2 \sigma_{\rm{n}} d\mathbf{r}$. The capture rate $c_{\rm{n}}^{\rm{STM}}$ in s$^{-1}$ is obtained by dividing $C_{\rm{n}}^{\rm{STM}}$ by the elementary charge and assuming a neutral DB, i.e. $f_{\rm{DB}}^{*}=0$.

\end{appendices}
\newpage


\bibliography{DB_refs}{}
\bibliographystyle{unsrt}
 
\end{document}